\UseRawInputEncoding
\documentclass[10pt,conference,a4paper]{IEEEtran}
\usepackage{graphicx} 
\usepackage{amsmath}
\usepackage{float}
\usepackage{caption}
\usepackage{url}

\usepackage[boxed,ruled,vlined,linesnumbered,commentsnumbered, noresetcount]{algorithm2e}
\usepackage{algorithmic}

\usepackage{diagbox}
\usepackage{tikz}
\usetikzlibrary{shapes.geometric}
\usetikzlibrary{positioning}

\usepackage{subcaption}

\usepackage{soul}

\usepackage{tabularx}
\usepackage{ragged2e} 
\newcolumntype{L}{>{\RaggedRight}X} 
\usepackage{lipsum} 
\usepackage{amsmath}
\usepackage{amssymb}
\usepackage{comment}
\usepackage{makecell} 
\usepackage{multirow}

\usepackage{pifont}

\makeatletter
\newcommand\notsotiny{\@setfontsize\notsotiny\@vipt\@viipt}
\makeatother

\begin{document}


\title{Comparative Performance Evaluation of 5G-TSN Applications in Indoor Factory Environments}

\author{Kouros Zanbouri, Md. Noor-A-Rahim, Dirk Pesch\\
School of Computer Science and Information Technology\\ University College Cork, Ireland 
}


\maketitle

\begin{abstract}
While Time-Sensitive Networking (TSN) enhances the determinism, real-time capabilities, and reliability of Ethernet, future industrial networks will not only use wired but increasingly wireless communications. Wireless networks enable mobility, have lower costs, and are easier to deploy. However, for many industrial applications, wired connections remain the preferred choice, particularly those requiring strict latency bounds and ultra-reliable data flows, such as for controlling machinery or managing power electronics. The emergence of 5G, with its Ultra-Reliable Low-Latency Communication (URLLC) promises to enable high data rates, ultra-low latency, and minimal jitter, presenting a new opportunity for wireless industrial networks. However, as 5G networks include wired links from the base station towards the core network, a combination of 5G with time-sensitive networking is needed to guarantee stringent QoS requirements. In this paper, we evaluate 5G-TSN performance for different indoor factory applications and environments through simulations. Our findings demonstrate that 5G-TSN can address latency-sensitive scenarios in indoor factory environments.

\end{abstract}



\section{Introduction}

The vision of future smart manufacturing and Industry 4.0 is driving the development of flexible, reconfigurable manufacturing environments that adapt in real time to market demands. Such environments will make use of mobile sensors, robotics, automated guided vehicles (AGVs), and drones to enable reconfiguration for manufacturing customized goods. As wired automation technologies lack the flexibility to meet reconfiguration requirements, wireless connectivity is increasingly introduced into manufacturing environments.

Industrial automation has long relied on proprietary field bus networks. IEEE~802.1 Time-Sensitive Networking (TSN) enables the transmission of data traffic for time-sensitive and mission-critical applications over a standard bridged Ethernet network, which is shared with various applications with different Quality of Service (QoS) requirements. Through the use of diverse queuing and shaping algorithms and the allocation of resources specifically for critical traffic, TSN guarantees that critical data traffic experiences no congestion loss. This allows TSN to ensure a worst-case, bounded end-to-end latency for time-critical data. Furthermore, TSN provides ultra-reliable data transmission by guarding against transmission failures using packet-level reliability mechanisms. Several standardisation groups are creating TSN profiles for certain application domains in an effort to simplify the deployment of TSN networks, such as the IEC/IEEE 60802 TSN Profile for Industrial Automation \cite{iec_ieee_60802}.

Currently, wireless communication in industrial settings leans towards specialized tasks and IT-related functions, typically for non-critical applications. However, expanding its reach to critical processes unlocks significant value. The most impactful benefit is enabling reliable connections for mobile systems like robots, AGVs, drones, and personnel. Seamless movement and constant connectivity are essential for reliable operation, particularly those involving AGVs or mobile control panels with safety features \cite{9855453}. 5G is a leading candidate for Wireless TSN in Industry 4.0 applications due to its design for real-time communication 
While prior studies have shown 5G's capability to achieve low latency under specific configurations \cite{8329620}, 
5G-TSN integration faces challenges like time synchronization and resource management.
Existing research on 5G-TSN systems has also not yet considered realistic wireless factory environments, rendering their results less practical. To the best of our knowledge, this work is the first to consider a wireless factory environment as outlined in standard specifications. The presented simulation-based investigation provides a more accurate evaluation by implementing an indoor factory profile within a 5G-TSN network, offering the following key contributions: 
\begin{itemize}
    \item Implementation of an Indoor Factory Profile based on 3GPP TR 38.901 within a 5G-TSN network model, representing a real-world wireless indoor factory environment

    \item Performance analysis using different use cases and factory environments.

    \item Study of the impact of different Indoor Factory profiles, including the effects of mobility and distance on network performance.
\end{itemize}

The remainder of this paper is organized as follows. In Section \ref{sec:Preliminaries}, we discuss the fundamental concepts of TSN and 5G. In Section \ref{sec:Model&Scenarios}, we explore the specific characteristics of indoor factory environments, along with selected use cases of 5G-TSN. Section \ref{sec:Simulation} details the experimental results and performance evaluation. Finally, Section \ref{sec:Conclusion} summarizes the key findings and discusses potential future research directions.

\section{Preliminaries} \label{sec:Preliminaries}

\subsection{Time-Sensitive Networking}
The IEEE~802.1 TSN standards expand standard Ethernet capabilities by introducing features that ensure timely delivery, reliability, and efficient management of data. 
The foundational elements upon which TSN is constructed are time-synchronization, high availability and ultra-reliability, bounded low latency, and resource management, which will be elaborated upon in the following.
\subsubsection{Time-synchronization}
Accurate time synchronization is essential for TSN in ensuring consistent and predictable operation. The IEEE 802.1AS standard focuses on time synchronization by introducing a generalized Precision Time Protocol (gPTP) profile. This ensures precise time synchronization across a network using the IEEE 1588v2 protocol \cite{8412460}, down to the microsecond level. 
\subsubsection{High availability and ultra-reliability}
TSN must manage congestion and packet loss induced retransmissions for highly time-sensitive and safety-critical applications. To achieve this, the IEEE 802.1Qca and IEEE 802.1CB standards provide redundancy mechanisms for TSN. IEEE 802.1CB frame replication and elimination for reliability (FRER) improves reliability by sending packet duplicates over multiple paths. The IEEE~802.1Qca manages multiple paths in bridged networks, ensuring reliable data delivery. It controls path selection, reserves network resources, and coordinates data transmission for efficient communication \cite{electronics10010058}. IEEE~802.1Qci prioritizes time-sensitive traffic by filtering and policing QoS in individual data streams \cite{9307422}.
\subsubsection{Bounded low latency}
TSN networks prioritize time-sensitive packets using mechanisms like traffic scheduling, shaping, and preemption, defined in standards IEEE~802.1Qav, 802.1Qbv, 802.1Qbu, 802.1Qch, and 802.1Qcr \cite{4395377, 10.1145/2997465.2997470}. TSN mechanisms can handle different traffic types with varying QoS requirements in multi-hop networks. IEEE~802.1Qbv prioritizes time-critical traffic using time-triggered scheduling, while IEEE 802.1Qav handles less strict latency requirements using a credit-based approach for bandwidth allocation. In addition, IEEE~802.1Qbu and 802.3br allow high-priority (express) frames to interrupt lower-priority ones, improving network efficiency and reducing latency \cite{8714953}.
\subsubsection{Resource management}
To achieve low latency and high reliability, careful configuration and management of TSN devices and resources are essential, as defined in IEEE 802.1Q. IEEE~802.1Qcc defines three TSN configuration models: centralized, distributed, and hybrid. The standard also provides protocols for TSN network management and setup. The IEEE~802.1Qat Stream Reservation Protocol (SRP) ensures timely delivery of critical data by reserving network resources and establishing traffic schedules between devices, determined by the specific latency and bandwidth requirements of each data stream.
\subsection{5G and wireless TSN}
5G is currently the leading wireless technology for supporting real-time applications. 5G's ultra-reliable and low-latency communication (URLLC) capabilities have been shown to achieve latencies of 1 millisecond and cycle times of 2 to 3 milliseconds under specific conditions, both in simulations and practical testing. Recent advancements in 5G technology, spearheaded by the 3GPP, have significantly improved data speeds and reduced latency \cite{abdullah2021enhanced}.
However, wireless networks are inherently unpredictable due to environmental factors and communication parameters, leading to higher error rates, packet loss, latency, and jitter compared to wired networks. While TSN excels at managing latency and jitter in wired environments through bandwidth reservation and scheduling, its application to wireless networks requires careful consideration of the dynamic nature of wireless channels, including fluctuating data rates, error rates, and packet loss \cite{zanbouri2023comprehensive}. The most promising approach for incorporating TSN capabilities into 5G networks, as outlined by the 3GPP, involves seamlessly integrating the 5G infrastructure as a logical bridge for TSN services \cite{9921731}. 
Understanding the impact of wireless channels when integrating TSN into 5G networks is crucial for effectively managing TSN performance requirements. The following section looks into the specific characteristics of indoor factory environments along with the use cases of 5G-TSN we studied in this paper.

\section{System Model and Scenarios} \label{sec:Model&Scenarios}
\subsection{Indoor Factory Channel Modelling}
\begin{figure}
    \centering
    \includegraphics[width=0.7\linewidth]{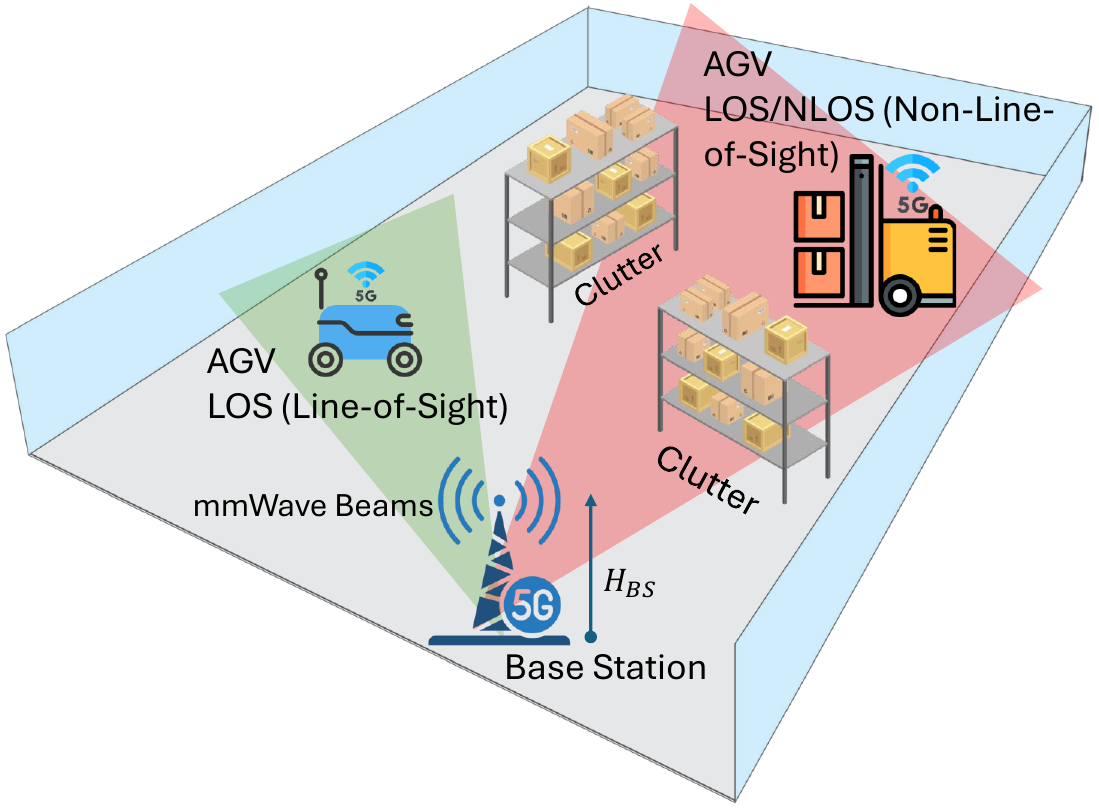}
    \caption{5G-TSN Indoor Factory Environment}
    \label{fig:5G-TSN-Overview}
\end{figure}

As wireless channels are significantly affected by their surroundings, key characteristics of the wireless system's deployment environment need to be captured when creating a model for a wireless channel. The 3rd Generation Partnership Project (3GPP) have detailed channel models for the 5G New Radio (NR) system across a wide range of environments in report 3GPP TR 38.901 \cite{3GPP_TR_38_901, 10118661}. Among others, the report outlines the indoor factory scenarios, which include factory buildings of various sizes and densities with different levels of clutter such as machinery, assembly lines, and storage shelves. As outlined in the report, the following indoor factory (InF) profiles are defined: InF-SL (sparse clutter, low base station), InF-DL (dense clutter, low base station), InF-SH (sparse clutter, high base station), InF-DH (dense clutter, high base station), and InF-HH (high transmitter, high receiver).
Fig. \ref{fig:5G-TSN-Overview} shows a 5G-TSN network in an industrial environment illustrating the elements of the indoor factory profile with both LOS and NLOS scenarios.

The path loss and line-of-sight (LOS) probability models for the 3GPP TR 38.901 channel models consider factors like antenna height and density of objects. While the indoor office scenarios typically involve objects like people, light partitions, desks, and chairs with limited heights, the InF scenario features larger machinery that can significantly block signals. In the InF, the ceiling height also varies widely, from 5 to 25 meters. To account for these differences, the 3GPP report uses terms like ``clutter-embedded'' and ``clutter-elevated'' to describe the height of the user terminal (UT) and base station (BS). For path loss, the model considers one LOS and four NLOS (Non-line-of-sight ) scenarios to reflect the different blocking conditions that can occur in industrial environments. The InF LOS path loss is calculated using Eq. \ref{eq:(1)}:
\begin{equation}
\label{eq:(1)}
\begin{aligned}
PL_{\text{LOS}} = 31.84 + 21.5 \cdot \log_{10} \left({d_{3D}} \right) + 19 \log_{10} \left({f_c} \right) \\ 
\sigma_{SF} = 4
\end{aligned}
\end{equation}
where \(d_{3D}\) is the 3D distance between transmitter and receiver (in meters), constrained by \(1 \leq d_{3D} \leq 600\) meters; \(f_c\) denotes the carrier frequency (in GHz); and \(sigma_{SF}\) is the shadow fading standard deviation (in dB).
NLOS path loss of the InF-SL is calculated using Eq. \ref{eq:InF-SL}:
\begin{equation}
\label{eq:InF-SL}
\begin{aligned}
PL &= 33 + 25.5 \log_{10} \left({d_{3D}} \right) + 20 \log_{10} \left({f_c} \right) \\
PL_{\text{NLOS}} &= \text{max}\, (PL, PL_{LOS}), \sigma_{SF} = 5.7
\end{aligned}
\end{equation}
For InF-DL, the NLOS path loss is calculated using Eq. \ref{eq:InF-DL}:
\begin{equation}
\label{eq:InF-DL}
\begin{aligned}
PL &= 18.6 + 35.7 \log_{10}(d_{3D}) + 20 \log_{10}(f_{c}) \\
PL_{\text{NLOS}} &= \text{max}\, (PL, PL_{\text{LOS}}, PL_{\text{InF-SL}}), \sigma_{SF} = 7.2
\end{aligned}
\end{equation}
InF-SH has a different path loss calculation for NLOS, which uses Eq. \ref{eq:InF-SH}:
\begin{equation}
\label{eq:InF-SH}
\begin{aligned}
PL &= 32.4 + 23.0 \log_{10}(d_{3D}) + 20 \log_{10}(f_{c}) \\
PL_{\text{NLOS}} &= \text{max}\, (PL, PL_{\text{LOS}}) , \sigma_{SF} = 5.9
\end{aligned}
\end{equation}
Finally, for InF-DH, the NLOS is calculated with Eq. \ref{eq:InF-DH}:
\begin{equation}
\label{eq:InF-DH}
\begin{aligned}
PL &= 33.63 + 21.9 \log_{10}(d_{3D}) + 20 \log_{10}(f_{c}) \\
PL_{\text{NLOS}} &= \text{max}\, (PL, PL_{\text{LOS}}), \sigma_{SF} = 4
\end{aligned}
\end{equation}
The LOS probability for InF-SL, InF-SH, InF-DL, and InF-DH profiles can be calculated using Eqs. \ref{eq:los_prob}, \ref{eq:k_subsec}, \ref{eq:los_hh}:
\begin{equation}
\text{Pr}_{\text{LOS, subsec}}(d_{2D}) = \exp\left(-\frac{d_{2D}}{k_{\text{subsec}}}\right)
\label{eq:los_prob}
\end{equation}
where \begin{equation}
k_{\text{subsec}} = 
\begin{cases} 
    \frac{-d_{\text{clutter}}}{\ln(1 - r)} & \text{for InF-SL and InF-DL} \\
    \frac{-d_{\text{clutter}}}{\ln(1 - r)} \cdot \frac{h_{\text{BS}} - h_{\text{UT}}}{h_{c} - h_{\text{UT}}} & \text{for InF-SH and InF-DH}
\end{cases}
\label{eq:k_subsec}
\end{equation}

The parameters $d_{\text{clutter}}$, $r$, and $h_c$ are specified in Table 7.2-4 of 3GPP TR 38.901. For InF-HH, the LOS probability is \begin{equation} \text{Pr}_{\text{LOS}} = 1 \label{eq:los_hh} \end{equation}

\subsection{Scenarios/Use Cases}
An AGV (Automated guided vehicle) is a versatile platform taking advantage of 5G connectivity in multiple contexts. AGVs have originally been employed to transport materials within factories. In recent years, AGVs and AMRs (Autonomous mobile robots) have become increasingly common in flexible manufacturing systems and collaborative robotics, where they automate tasks that were once done by human workers. A 5G-enabled AGV can enhance safety and monitoring activities by autonomously inspecting factory areas for potential hazards and aiding in overall factory operations. The integration of AGVs and AMRs into industrial logistics, particularly within indoor factory environments, has been a subject of significant interest in industry. Companies such as Amazon \cite{amazon}, DHL, Foxconn, Tesla, Samsung have adopted these technologies to streamline their operations and enhance operational efficiency. Additionally, researchers have conducted extensive studies to explore the potential benefits, challenges, and best practices associated with 5G-based AGV/AMR implementation in manufacturing applications \cite{9247159}. 

\begin{figure}
    \centering
    \includegraphics[width=0.8\linewidth]{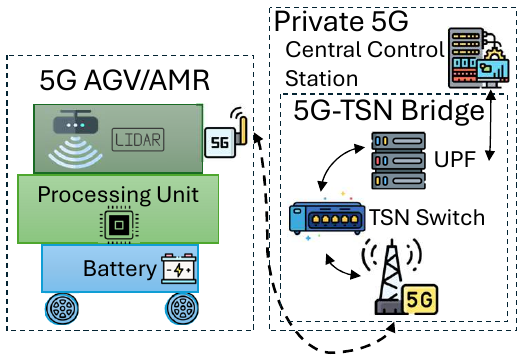}
    \caption{5G-Based MiR architecture in 5G-TSN network}
    \label{fig:5G-TSN-MIR}
\end{figure}

To assess the effectiveness of 5G-TSN in industrial settings, we consider a scenario with a mobile industrial robot (MiR) similar to MiR250 or MiR1200 pallet jack \cite{mir250} equipped with 5G connectivity. The MiR comprises a main battery, a digital camera equipped with Light Detection and Ranging (LiDAR) sensors connected to a video processing workstation unit, and a 5G industrial router. The 5G router connects directly to a standalone 5G base station, forming a private 5G network. 
The MiR's Central Control (CC) station is also part of this private network, offering mobile edge computing (MEC) capabilities and facilitating a connection to the MiR's web interface through a robotic operating system bridge. This setup also enables remote desktop access to the AGV and access to its LiDAR streaming port. Fig.~\ref{fig:5G-TSN-MIR} illustrates the components and connectivity of a 5G-based MiR with the 5G network and command and control station. 

In our 5G-TSN network use case, we have categorized traffic into three types: Network Control (NC), Video, and Best Effort (BE). NC and Video are both TSN traffic streams, with NC having the highest priority, video having medium priority, and BE, a non-TSN traffic type, having the lowest priority. NC traffic is generated periodically, with a time interval ranging between 50 ms and 1 second, and packet size variation between 50 and 500 bytes. The NC transmits time-critical data related to the AGV's navigation, control, and safety systems. This includes information such as position, velocity, obstacle detection, and commands from the central control system. It is characterized by regular, predictable intervals to ensure real-time network performance.
The video stream traffic is also periodic, reflecting its real-time nature and continuous data flow requirements. The video transmits video data from the AGV's camera. This could be used for tasks like object recognition, visual inspection, and remote monitoring. The packet size for this type of traffic fluctuates between 1000 and 1500 bytes, representing the typical variation in video frame sizes due to dynamic scene content. The periodicity ensures a consistent stream of data, which is essential for maintaining video quality and reducing latency. In contrast, BE traffic is sporadic, characterized by irregular intervals that range from 500 ms to 2 seconds. The packet size for this traffic can vary significantly, from as small as 30 bytes to as large as 1500 bytes. The BE traffic transmits data that is not critical for the AGV's immediate operation, such as telemetry, diagnostics, or software updates. This traffic occurs without a fixed or predictable pattern, often experiencing periods of inactivity followed by bursts of high activity. As a result, it has the lowest priority in time-critical systems, and its delivery is subject to network availability, making it suitable for non-essential, background tasks where timing is less critical.
This classification and characterization of traffic types allow for efficient network resource allocation in time-sensitive environments, ensuring that NC and video data are prioritized, while best-effort traffic is handled opportunistically.

\section{Simulation Results and Discussion} \label{sec:Simulation}
For our simulation, we employed the OMNET++ 6.0.3 simulation environment \cite{10.4108/ICST.SIMUTOOLS2008.3027}, incorporating the iNet 4.5.2 \cite{Mészáros2019}, Simu5G 1.2.2 \cite{Simu5G}, and 5GTQ \cite{10333533} frameworks. The specific parameters employed in our simulation are detailed in Table~\ref{tab:simulation}.
\begin{table}
    \centering
    \begin{tabular}{|c|c|}
    \hline
       Parameter & Value \\ \hline
        No. of UEs & 1\\ \hline
        No. of gNBs & 1 \\ \hline
        gNB Tx power & 23 dBm \\ \hline
        Target Bler  & 0.01 \\ \hline
        Physical environment & FlatGround \\ \hline
        Carrier frequency & 5.9 GHz \\ \hline
        Numerology index & 4 \\ \hline
        Channel model & Indoor Factory\\ \hline
        InF profiles & InF-SL, InF-DL, InF-SH, InF-DH\\ \hline
        Mobility model & Random Waypoint Mobility\\ \hline
        Mobility speed & 0.2-1.5 mps \\ \hline
    \end{tabular} 
    \caption{5G Configuration settings for the simulated environment}
    \label{tab:simulation}
\end{table}
In the simulated 5G-TSN network, each flow of data has a specific quality of service (QoS) profile. This profile details the type of resources needed, the priority level, how much delay is acceptable, the maximum error rate, and other important factors. These profiles are identified by a unique number called 5QI. The 5QI numbers are grouped into three categories based on the type of resources they need: guaranteed bit rate (GBR), non-guaranteed bit rate (Non-GBR), and delay-critical guaranteed bit rate (DC-GBR). TSN uses the Priority Code Point (PCP) field to prioritize different types of data. In this study, PCP values range from 0 to 7, with 0 being the lowest priority and 7 being the highest.

\begin{figure}
    \centering
    \includegraphics[width=0.9\linewidth]{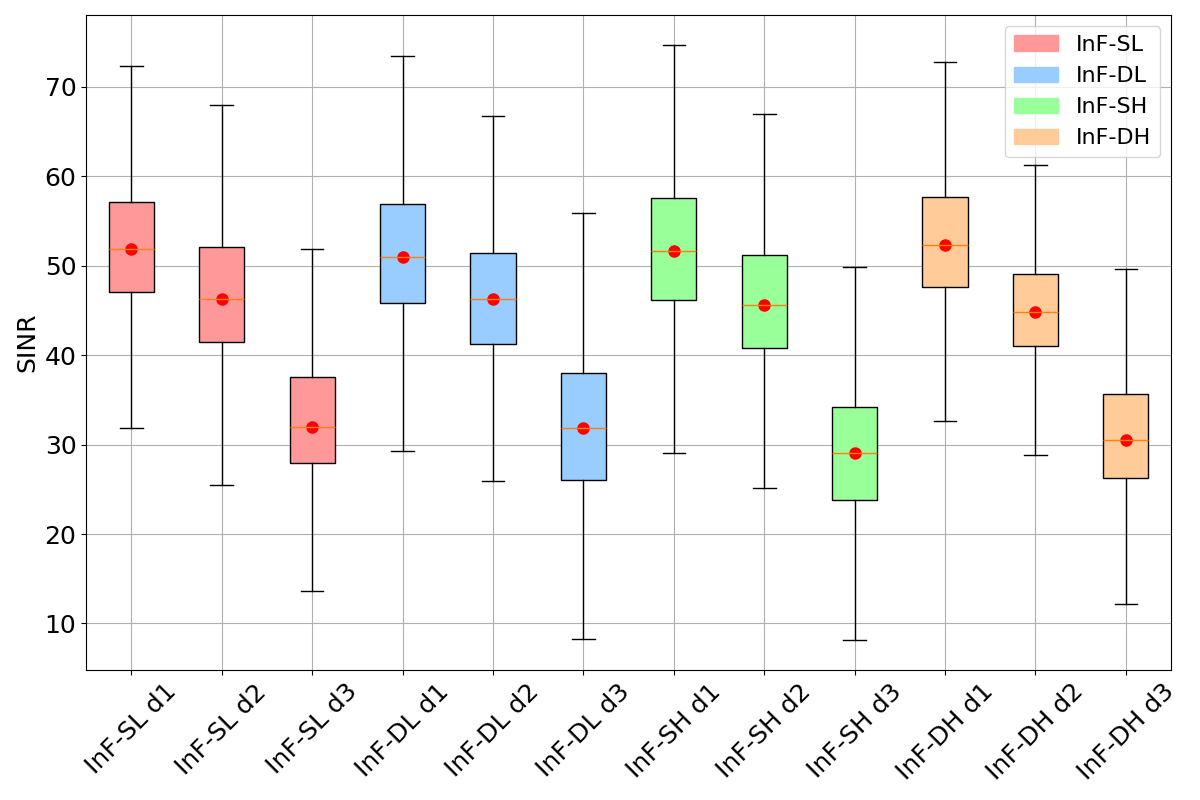}
    \caption{SINR evaluation with different InF profile}
    \label{fig:sinr-profiles}
\end{figure}

To evaluate 5G-TSN in an industrial environment, we implemented the InF profiles into Simu5G. Based on the received power, the simulated receiver calculates the SINR and queries the Binder to identify which nodes are interfered with on the same resources. The Binder module stores node references and identifies interfering gNBs to calculate UE inter-cell interference. The receiver then uses Block Error Rate (BLER) curves to estimate the reception probability for each resource block involved in the transmission. This process enables the conversion of SINR and the transmission format into a probability of correctly receiving the entire MAC PDU.
Fig. \ref{fig:sinr-profiles}  evaluates the SINR performance of different InF profiles in the downlink, each using its default parameter values. The experimental findings demonstrate a clear disparity in SINR performance between the InF-DL and InF-SH profiles. The lowest SINR was observed in InF-DL profile, indicating a more challenging radio environment for downlink transmissions. Conversely, the the highest SINR values were achieved in InF-SH profile, suggesting a more favourable radio environment for short-range, high-rate transmissions.

To evaluate each indoor factory profile in our use-case scenario, we divided the distance between the UE and the gNB into three distinct sections: d1, d2, and d3. These divisions represent increasing radial distances from the gNB. The d1 section encompasses a radius of 85 meters from the gNB, representing the closest proximity for UEs. The d2 section extends to a radius of 170 meters, while the d3 section covers a radius of 255 meters from the gNB. These defined distance segments allow for a systematic analysis of performance metrics, accounting for the variation in signal strength and network performance as the distance from the gNB increases. By segmenting the use-case scenario into these three distance ranges, we aim to better understand the impact of distance on communication reliability and efficiency in an indoor factory setting.

\begin{figure}
    \centering
    \includegraphics[width=0.95\linewidth]{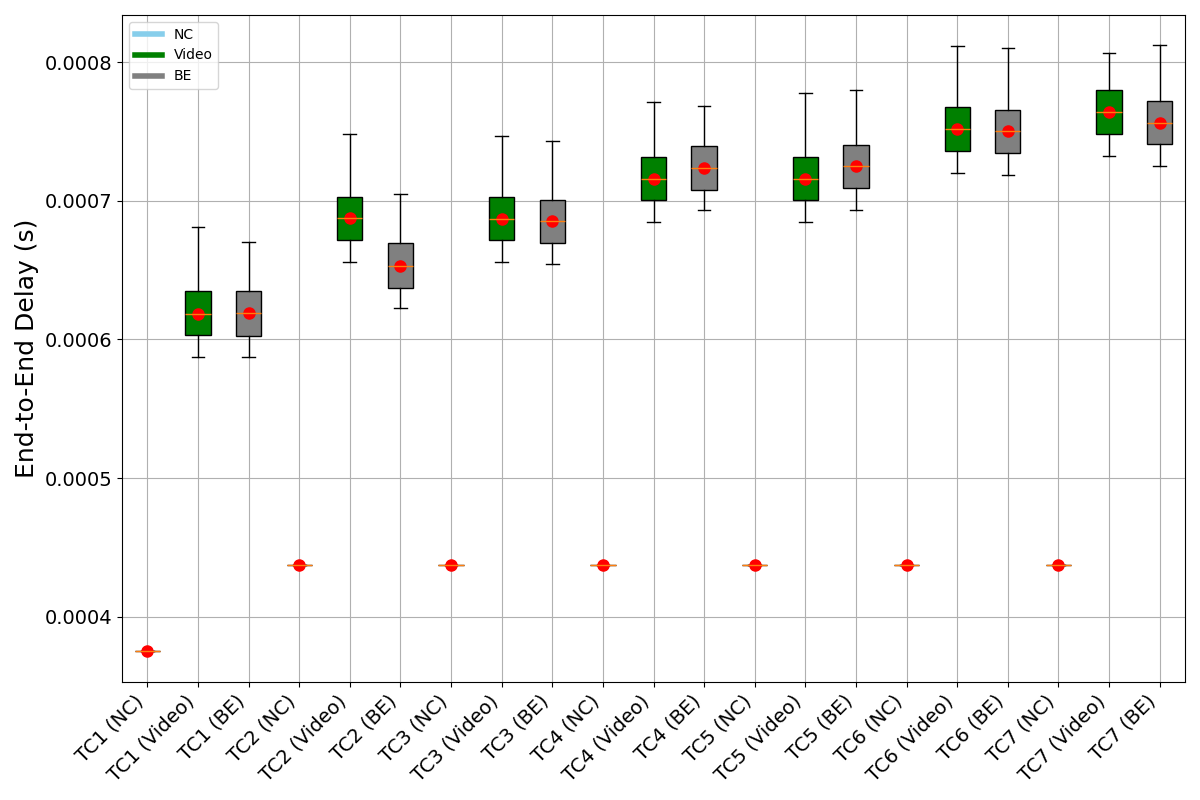}
    \caption{End-to-End delay with varying data rate}
    \label{fig:E2E-DataRate}
\end{figure}

Fig. \ref{fig:E2E-DataRate} illustrates the impact of varying data rates on end-to-end latency for each data stream and with node speeds ranging from $0.2$ m/s to $1.5$ m/s, Random Waypoint Mobility and maximum distance constraint of 250-meter from the gNB. The end-to-end latency tracks how long it takes a data packet to travel from its starting point (the source application) to its final destination (the destination application).
Table \ref{tab:E2E-Parameters} summarises the test case parameters utilized in the simulation. The analysis of the box plots reveals significant differences in the performance stability of different traffic classes across multiple test cases (TC1 to TC7). The NC traffic exhibits a high degree of stability and consistency, with minimal variation in both the interquartile range (IQR) and whiskers across all test cases, indicating that this traffic class is reliably managed under varying conditions. In contrast, the Video and BE traffic classes demonstrate greater variability, with more pronounced changes in the whiskers and the presence of outliers across different test cases. This suggests that these traffic classes are more susceptible to performance fluctuations based on test case conditions. The whiskers for Video and BE traffic extend further, highlighting a wider distribution of values and indicating instability in comparison to the NC traffic.
\begin{table*}[]
\centering
\begin{tabular}{|l|l|l|l|l|l|l|}
\hline
Test-Case & \begin{tabular}[c]{@{}l@{}}NC\\ Rate (Kbps)\end{tabular} & \begin{tabular}[c]{@{}l@{}}Video\\ Rate (Kbps)\end{tabular} & \begin{tabular}[c]{@{}l@{}}BE\\ Rate (Kbps)\end{tabular} & \begin{tabular}[c]{@{}l@{}}NC\\ Packetlength (B),\\ Interarrival (ms)\end{tabular} & \begin{tabular}[c]{@{}l@{}}Video\\ Packetlength (B),\\ Interarrival (ms)\end{tabular} & \begin{tabular}[c]{@{}l@{}}BE\\ Packetlength (B),\\ Interarrival (ms)\end{tabular} \\ \hline
TC1       & 48                                                       & 114.2                                                       & 11.4                                                     & 300, 50                                                                            & 1000, 70                                                                               & 1000, 700                                                                           \\ \hline
TC2       & 68.32                                                    & 138.62                                                      & 12.67                                                    & 427, 50                                                                            & 1213, 70                                                                               & 1109, 700                                                                           \\ \hline
TC3       & 56.93                                                    & 138.62                                                      & 19.34                                                    & 427, 60                                                                            & 1213, 70                                                                               & 1209, 500                                                                           \\ \hline
TC4       & 64.66                                                    & 148.91                                                      & 21.28                                                    & 485, 60                                                                            & 1303, 70                                                                               & 1330, 500                                                                           \\ \hline
TC5       & 55.42                                                    & 148.91                                                      & 19.34                                                    & 485, 70                                                                            & 1303, 70                                                                               & 1330, 550                                                                           \\ \hline
TC6       & 79.68                                                    & 161.48                                                      & 20.49                                                    & 498, 50                                                                            & 1413, 70                                                                               & 1409, 550                                                                           \\ \hline
TC7       & 72.43                                                    & 193.73                                                      & 19.05                                                    & 498, 55                                                                            & 1453, 60                                                                               & 1429, 600                                                                           \\ \hline
\end{tabular}
    \caption{Parameter settings used in the simulation to evaluate end-to-end delay across different traffic scenarios for InF-SL. NC, Video, and BE rates are given in kbps, with packet lengths (B) and interarrival times (ms) specified for each test case.}
    \label{tab:E2E-Parameters}
\end{table*}

\begin{figure}
    \centering
    \includegraphics[width=0.9\linewidth]{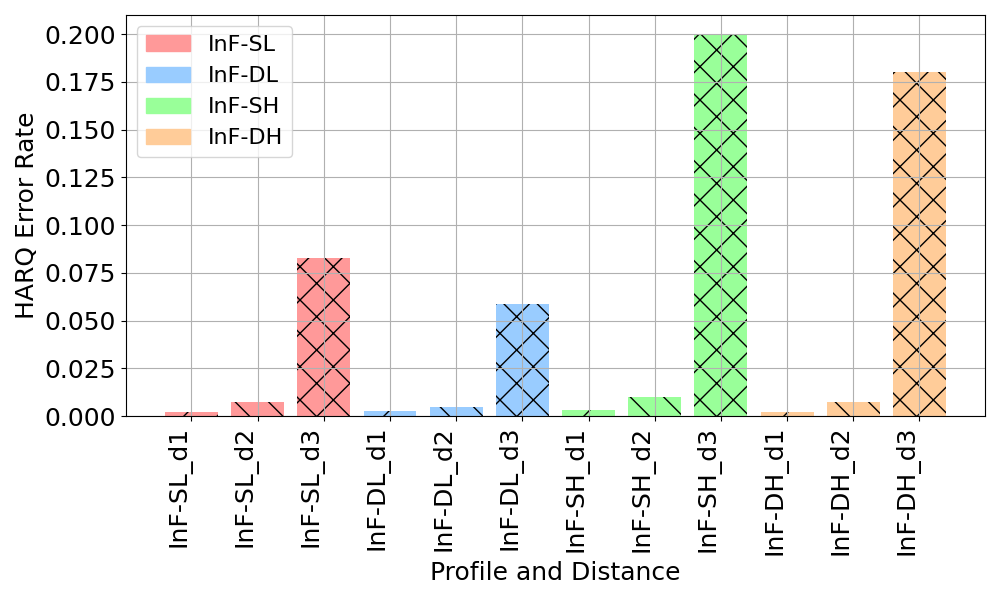}
    \caption{HARQ error rate with different profiles and distances}
    \label{fig:harqerror}
\end{figure}

Fig. \ref{fig:harqerror} illustrates the behaviour of the radio interface's Hybrid Automatic Repeat reQuest (HARQ) under varying distance conditions and indoor factory profiles. It compares the performance of time-critical and non-critical data transmission in relation to distance within the indoor factory environment. The HARQ error rate, calculated as the ratio of failed transmissions to total transmissions, provides insight into the reliability of these transmissions under different scenarios.
This analysis demonstrates that as distance increases, the HARQ error rate tends to increase across all profiles, but the magnitude of the increase depends on the specific profile. InF-SL and InF-SH suffer the most as distance increases. On the other hand, InF-DL and InF-DH show better resilience to increasing distance but still encounter a rise in error rates beyond d2.
This trend suggests that in indoor factory environments, profiles with higher latency tolerance (like InF-DL and InF-DH) may accommodate longer-range communication, while profiles designed for lower latency or higher throughput need careful design in terms of distance to avoid performance degradation. 
\begin{figure}
  \centering  
  \begin{subfigure}{0.485\textwidth}  
    \includegraphics[width=\linewidth]{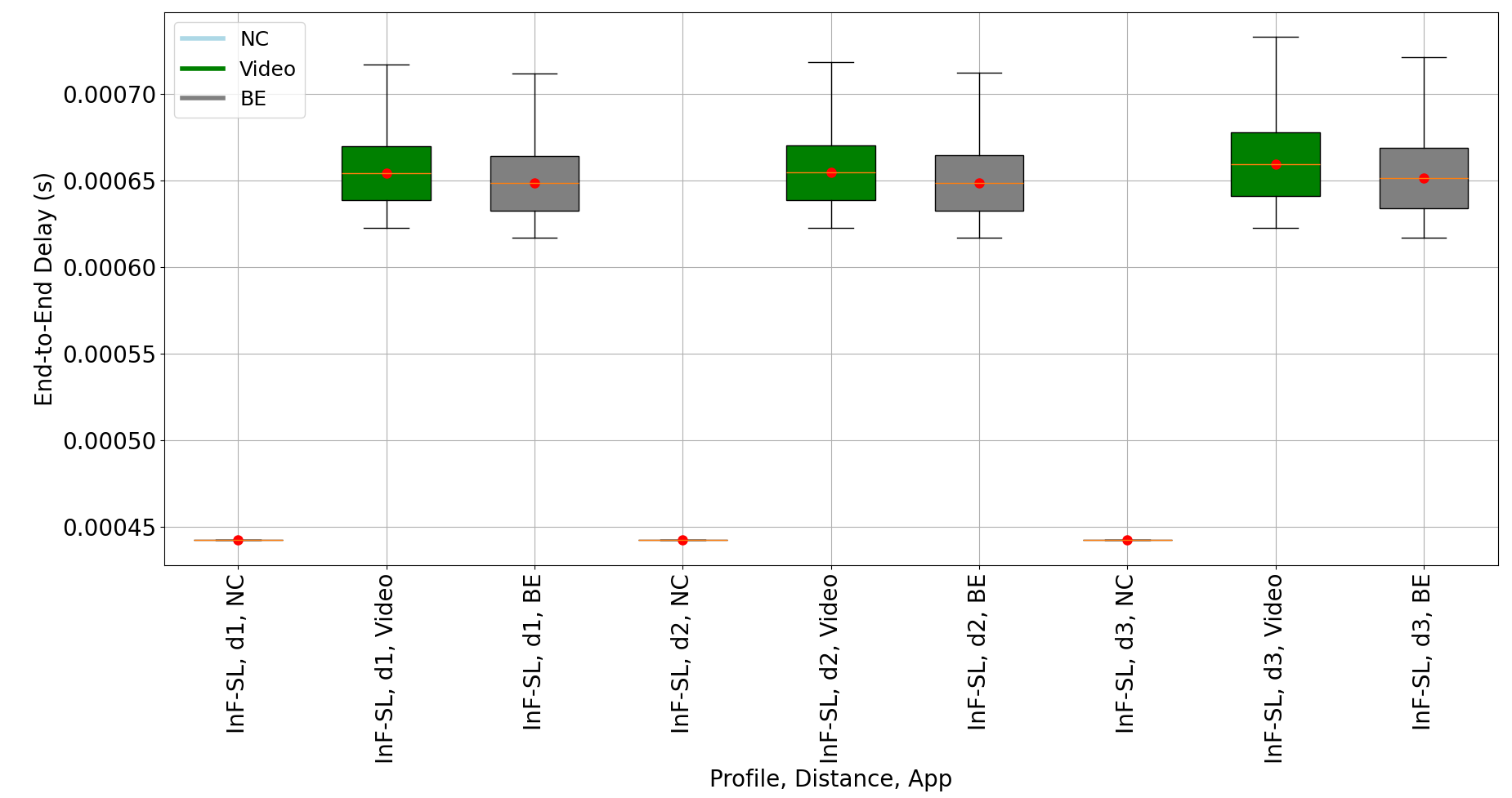}
    \caption{InF-SL profile across different distances}  
    \label{fig:6a}
  \end{subfigure}
  \hspace{0.05\textwidth}  
  \begin{subfigure}{0.485\textwidth}  
    \includegraphics[width=\linewidth]{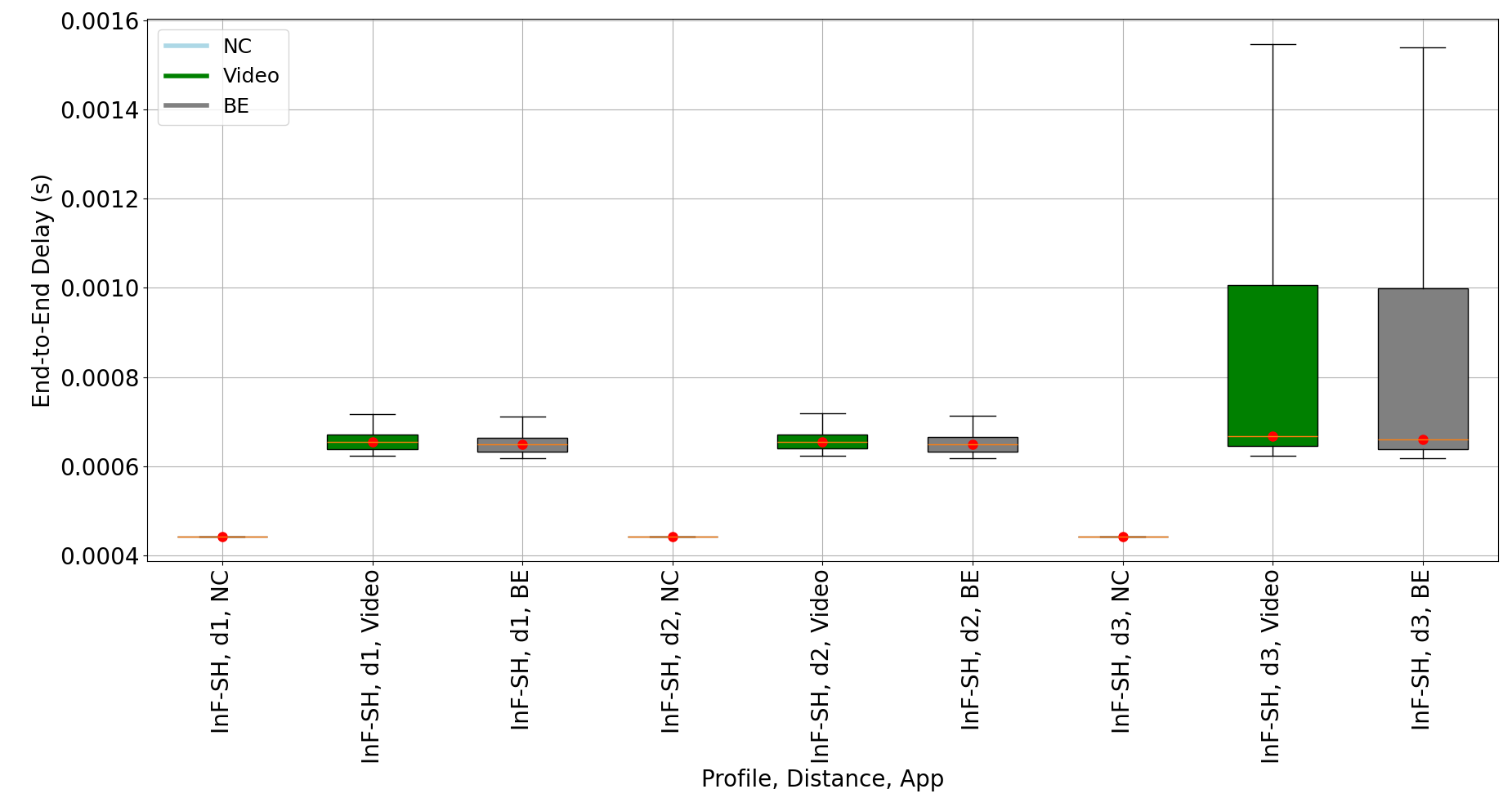}
    \caption{InF-SH profile across different distances}  
    \label{fig:6b}
  \end{subfigure}
  \caption{End-to-End delay with different profiles and distances}  
  \label{fig:6-end-to-end-distance}
\end{figure}
Fig. \ref{fig:6-end-to-end-distance} presents the evaluation of end-to-end delay for two profiles, InF-SL and InF-SH, across varying distances. As illustrated in Fig. \ref{fig:harqerror}, increasing the distance leads to a rise in HARQ error rate, which results in more retransmissions. Based on the HARQ error rate and end-to-end latency figures, the performance of InF-SL and InF-SH profiles demonstrates notable distinctions in reliability and latency as the distance increases. For the HARQ error rate, both profiles exhibit an increasing trend as the distance between UE and gNB progresses from d1 to d3, with InF-SH showing a higher error rate at longer distances compared to InF-SL. This indicates that InF-SH is more susceptible to packet retransmission errors. In terms of end-to-end latency, InF-SH demonstrates a greater variation, for traffic streams like video and BE, at longer distances (d3). Conversely, InF-SL maintains relatively consistent latency performance across all distances, albeit with slightly higher error rates than profiles such as InF-DL or InF-DH. This trade-off between reliability (HARQ error rate) and latency underscores the challenges in optimizing network performance for these profiles, particularly in scenarios with longer distances.

\section{Conclusion} \label{sec:Conclusion}
The move towards Industry 4.0 necessitates that factories replace their current industrial networks, which rely on Ethernet and other wired technologies, with wireless networks. This transition is essential for achieving the desired flexibility and adaptability in manufacturing environments. While some companies have begun to experiment with 5G technology, deployments are still at an early stage of development. In this study, we have simulated a 5G-TSN network deployed in an indoor factory to assess its suitability for industrial applications. We conducted simulations of 5G-TSN across various factory environments to understand its behaviour under different wireless channel profiles. Our results indicate that 5G-TSN can effectively handle latency-critical applications in indoor factories in a controlled environment, suggesting its potential for significant performance improvements. Since no prior work has been done in this specific context, our research serves as a baseline for evaluating 5G-TSN in factory scenarios. However, we plan to extend this work by testing more realistic scenarios with real-world parameters, such as varying numbers of UEs, gNB, varying network load, and more dynamic environmental interference. These tests will allow us to assess the scalability, reliability, and adaptability of the proposed approach in more complex and dynamic indoor factory environments in a more comprehensive manner. Based on the findings of this study, future work could also focus on real-world implementation and testing of 5G-TSN in industrial environments to validate the simulation results and identify any challenges or limitations. Additionally, research could explore the integration of 5G-TSN with existing factory automation systems and infrastructure to ensure seamless compatibility and maximize its benefits.

\section*{Acknowledgments}
This publication has emanated from research conducted with the financial support of Research Ireland under Grant number 13/RC/2077\_P2. 

\bibliography{references}
\bibliographystyle{IEEEtran}

\end{document}